\documentclass[12pt,a4paper]{article}
\usepackage{amsfonts,latexsym}
\usepackage{graphicx,color}
\usepackage{dcolumn}
\usepackage{graphicx}
\usepackage{amsmath}
\usepackage{amsfonts}
\usepackage{amssymb}

\renewcommand{\thefootnote}{\fnsymbol{footnote}}

\begin{document}

\vspace{12mm}

\begin{center}
{{{\Large {\bf Scalarized charged  black holes in the Einstein-Maxwell-Scalar theory with two U(1) fields}}}}\\[10mm]

{Yun Soo Myung$^a$\footnote{e-mail address: ysmyung@inje.ac.kr} and De-Cheng Zou$^{a,b}$\footnote{e-mail address: dczou@yzu.edu.cn}}\\[8mm]

{${}^a$Institute of Basic Sciences and Department  of Computer Simulation, Inje University Gimhae 50834, Korea\\[0pt] }

{${}^b$Center for Gravitation and Cosmology and College of Physical Science and Technology, Yangzhou University, Yangzhou 225009, China\\[0pt]}
\end{center}
\vspace{2mm}

\begin{abstract}
We investigate  scalarized charged  black holes  in  the Einstein-Maxwell-Scalar theory with two U(1) fields inspired  by the $N=4$ supergravity.
From  the onset of the  spontaneous scalarization (tachyonic instability of Reissner-Nordstr\"{o}m black hole),
these  black holes are classified   by the number of $n=0,1,2,\cdots$, where $n=0$ is called the fundamental black hole and $n=1,2,\cdots$ denote the $n$-excited black holes.
Adopting  radial perturbations, we show that   the $n=0$  black hole is stable against the $s(l=0)$-mode scalar perturbation, whereas the $n=1,2$ excited black holes are unstable.  This implies that  the $n=0$ black hole is  considered as an endpoint of  the Reissner-Nordstr\"{o}m black hole.

\end{abstract}
\vspace{5mm}

\vspace{1.5cm}

\hspace{11.5cm}
\newpage
\renewcommand{\thefootnote}{\arabic{footnote}}
\setcounter{footnote}{0}


\section{Introduction}
Recently, the inclusion of non-minimal scalar couplings with coupling parameter $\alpha$ has induced  the instability of Schwarzschild  black holes and thus, led to scalarized black holes~\cite{Doneva:2017bvd,Silva:2017uqg,Antoniou:2017acq}. This is known to be a phenomena of spontaneous scalarization, a way of providing black holes with scalar hair.
Also, non-minimal coupling to the Maxwell invariant [Einstein-Maxwell-scalar(EMS) theory]  has accommodated a phenomena of spontaneous scalarization of Reissner-Nodstr\"{o}m (RN) black holes~\cite{Herdeiro:2018wub}.
It is worth noting that the existence line  separating RN black holes from scalarized charged black holes is universal  in the sense that the various scalar couplings $\{f(\phi)\}$ to the Maxwell invariant  are identical in the linearized approximation~\cite{Fernandes:2019rez}.

On the other hand, an analysis of dilatonic   versus scalarized couplings  has shown that two have provided charged black holes with scalar hair with analytical and numerical forms, but the former does not accommodate RN black holes, whereas the latter has a smooth extremal scalarized black hole by considering dyonic RN black holes~\cite{Astefanesei:2019pfq}. This implies a comparative difference between dilatonic and  scalarized couplings  in the  EMS theory.

In this work, we wish to introduce the EMS theory with  different scalar couplings to two  U(1) field strengths  for realizing another spontaneous scalarization because the same coupling makes no difference.
This theory is inspired by the bosonic sector of $N=4$ supergravity which has admitted an analytically dilatonic black hole with a fixed scalar  including an extremal black hole.
There were many testing of stringy black holes with fixed scalars, being different  from minimally coupled (free) scalars. Such  testings have included  computation of the greybody factor for a propagating scalar around an extremal black holes: $\sigma^{\rm free}_s \to 4\pi$ and $\sigma^{\rm fixed}_s\to 4\pi\omega^2\to 0$ in the low-energy limit  ($\omega\to 0$), implying that a suppression of Hawking radiation occurred in the fixed scalar, compared to the free scalar~\cite{Kol:1996hf,Krasnitz:1997gn,Lee:1997xg}. For a fixed scalar, a scalar $\phi_{\rm fixed}^{\infty}$ at infinity   is independent of  the value of scalar $\phi(r_+)=\phi_{\rm fixed}^{0}(q)$ on the horizon~\cite{Ferrara:1995ih,Ferrara:1996um}. It is proposed that  a fixed scalar in the dilatonic black holes is similar to the scalar hair in the $n=0,~1,~2$ scalarized black holes.

Therefore, it is quite interesting to compare the fixed scalar in dilatonic black holes with the scalar hair in the sclarized black holes.
Introducing  radial perturbations, we wish to  show that   the $n=0$  black hole is stable against the $s(l=0)$-mode scalar perturbation,
while the $n=1,2$ excited black holes are unstable.

\section{Instability of RN black hole } \label{sec1}
First of all, we introduce the bosonic  action for $N=4$ supergravity~\cite{Kol:1996hf,Krasnitz:1997gn,Lee:1997xg}
\begin{equation}
S_{\rm N4}=\frac{1}{16 \pi}\int d^4 x\sqrt{-g}\Big[ R-2\partial_\mu \phi \partial^\mu \phi-e^{-2 \phi} F^2-e^{2 \phi} H^2\Big],\label{Act1}
\end{equation}
where $\phi$ plays the role of dilaton and   $F=dA$ and $H=dB$ are two U(1) field strengths. The RN-type black hole without scalar hair could not found  from (\ref{Act1}).
An analytic black hole solution is given by
\begin{equation} \label{n4bh}
ds^2_{\rm N4BH}=-\frac{1}{H_1H_2}dt^2+H_1H_2\Big(dr^2+r^2d\Omega^2_2\Big)
\end{equation}
and
\begin{equation}
e^{2\bar{\phi}}=\frac{H_2}{H_1},\quad \bar{F}=\frac{1}{\sqrt{2}}dH_1^{-1}\wedge dt,\quad  \bar{H}=\frac{1}{\sqrt{2}}dH_2^{-1}\wedge dt
\end{equation}
with two harmonic functions
\begin{equation}
H_1=1+\frac{\sqrt{2}Q}{r},\quad H_2=1+\frac{\sqrt{2}P}{r}.
\end{equation}
\begin{figure*}[t!]
   \centering
  \includegraphics{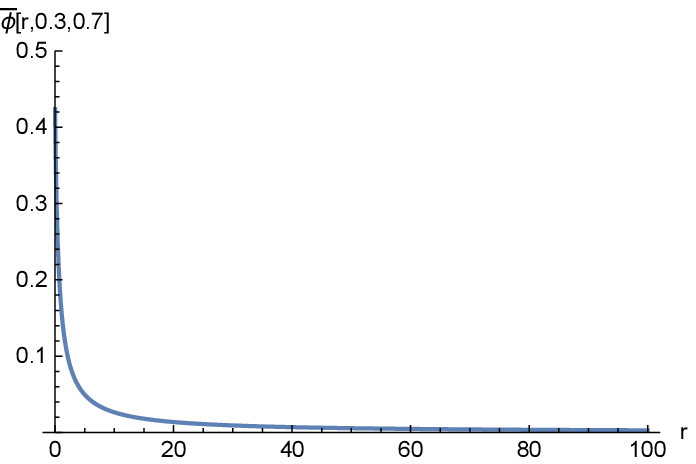}
  \hfill%
  \includegraphics{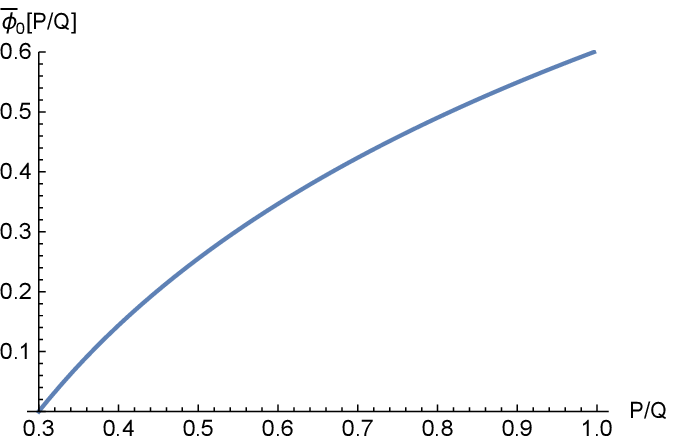}
\caption{(Left) The dilaton $\bar{\phi}(r,Q=0.3,P=0.7)$  as functions of $r\in [r_+=0,100]$. The dilaton takes a value of $\bar{\phi}= 0.42365$ on the horizon at $r=0$ and it vanishes asymptotically. (Right) The dilaton  $\bar{\phi}_0(P/Q)$ on the horizon at $r=r_+=0$ with $Q=0.3$, showing a fixed scalar.  }
\end{figure*}
The event horizon is located at $r_{+}=0$ and a fixed scalar $\bar{\phi}$ is defined as the special massless field whose value on the horizon is fixed by the U(1) charges $ Q$ and $P$, leading to $\lim_{r\to 0}\bar{\phi}=0.5\ln[\frac{P}{Q}]$. In case of $P=Q$(extremal black hole), one finds that $\bar{\phi}=0$. Fig. 1 shows that the dilaton is a fixed scalar, being similar to scalar hair. Considering the radial perturbations around (\ref{n4bh}), the linearized equation for $s$-mode dilaton $\delta \phi(t,r)=\tilde{\varphi}(r)e^{-i\omega t}$ is given by
\begin{equation}
\Big[\frac{1}{r^2}\frac{d}{dr}\Big(r^2 \frac{d}{dr}\Big)+\omega^2(H_1H_2)^2-\frac{4(P+Q)^2}{r^2(\sqrt{2}P+\sqrt{2}Q+2r)^2}\Big]\tilde{\varphi}(r)=0,
\end{equation}
which turned out to be stable because of $\omega>0$.

Now let us obtain  the action for the Einstein-Maxwell-Scalar theory with two U(1) fields (EMSN4 theory) induced by  $N=4$ supergravity   by replacing $2 \phi$ in the exponents with $\alpha \phi^2$ on (\ref{Act1})
\begin{equation}
S_{\rm EMSN4}=\frac{1}{16 \pi}\int d^4 x\sqrt{-g}\Big[ R-2\partial_\mu \phi \partial^\mu \phi-e^{-\alpha \phi^2} F^2-e^{\alpha \phi^2} H^2\Big],\label{Act2}
\end{equation}
where $\alpha$ is a scalar coupling parameter.

We  derive  the Einstein  equation from the action (\ref{Act2})
\begin{eqnarray}
 G_{\mu\nu}=2\partial _\mu \phi\partial _\nu \phi -(\partial \phi)^2g_{\mu\nu}+2T^{U(1)}_{\mu\nu} \label{equa1}
\end{eqnarray}
with $G_{\mu\nu}=R_{\mu\nu}-(R/2)g_{\mu\nu}$ and
\begin{equation}
T^{U(1)}_{\mu\nu}=e^{-\alpha \phi^2}\Big(F_{\mu\rho}F_{\nu}~^\rho-\frac{F^2}{4}g_{\mu\nu}\Big)+e^{\alpha \phi^2}\Big(H_{\mu\rho}H_{\nu}~^\rho-\frac{H^2}{4}g_{\mu\nu}\Big).
\end{equation}
Two  Maxwell equations take the forms
\begin{eqnarray} \label{M-eq}
&&\nabla^\mu F_{\mu\nu}-2\alpha \phi\nabla^{\mu} (\phi)F_{\mu\nu}=0,\label{M-eq1} \\
&&\nabla^\mu H_{\mu\nu}+2\alpha \phi\nabla^{\mu} (\phi)H_{\mu\nu}=0.
\end{eqnarray}
The scalar equation is given by
\begin{equation}
\square \phi +\frac{\alpha}{2}\Big(F^2e^{-\alpha \phi^2}-H^2e^{\alpha \phi^2}\Big) \phi=0 \label{s-equa}.
\end{equation}
First of all, we would like to mention the RN-type black hole solution without scalar hair
\begin{equation} \label{RN-bh}
ds^2_{\rm RN-type}=\bar{g}_{\mu\nu}dx^\mu dx^\nu=-f(r) dt^2+\frac{dr^2}{f(r)} +r^2d\Omega^2_2,\quad f(r)=1-\frac{2M}{r}+\frac{Q^2+P^2}{r^2}
\end{equation}
which  is obtained, irrespective of  any value of $\alpha$. Here, we have that  $\bar{\phi}=0$,  $\bar{A}_t=Q/r$, and $\bar{B}_t=P/r$.
Two horizons are determined as $r_\pm =M[1\pm \sqrt{1-(q^2+p^2)}]$ with $q=Q/M$ and $p=P/M$ by imposing $f(r)=0$. For $M=1$, one has $P=p$ and $Q=q$.
Hereafter, we consider only the region on and outside the outer horizon ($r\ge r_+$).   Further,
we would like to mention that the dyonic RN black hole takes the same form as  (\ref{RN-bh})~\cite{Astefanesei:2019pfq}.

Let us consider the perturbations around the background values
\begin{equation}
g_{\mu\nu}=\bar{g}_{\mu\nu}+h_{\mu\nu},\quad\phi=0+\delta \varphi, \quad F_{\mu\nu}=\bar{F}_{\mu\nu}+f_{\mu\nu},\quad H_{\mu\nu}=\bar{H}_{\mu\nu}+\tilde{f}_{\mu\nu},
\end{equation}
where
\begin{equation}
f_{\mu\nu} =\partial_\mu a_\nu-\partial_\nu a_\mu,\quad \tilde{f}_{\mu\nu}=\partial_\mu b_\nu-\partial_\nu b_\mu. \label{l-F}
\end{equation}
Now, we derive their linearized equations as
\begin{eqnarray}
&&\delta G_{\mu\nu}(h)=2 \delta T^{U(1)}_{\mu\nu}, \label{per-eq1}\\
&&\bar{\nabla}^\mu f_{\mu\nu}=0,\quad \bar{\nabla}^\mu \tilde{f}_{\mu\nu}=0,\label{per-eq2}\\
&&\Big[\bar{\square}+ \frac{\alpha(P^2-Q^2)}{r^4}\Big]\delta \varphi=0,\label{per-eq3}
\end{eqnarray}
where
\begin{eqnarray}
  \delta G_{\mu\nu} &=& \delta R_{\mu\nu}-\frac{1}{2} \bar{g}_{\mu\nu} \delta R -\frac{1}{2} \bar{R} h_{\mu\nu}, \label{l-G} \\
  \delta T^{U(1)}_{\mu\nu} &=&\bar{F}_{\nu}~^\rho f_{\mu\rho}+\bar{F}_{\mu}~^\rho f_{\nu\rho}-\bar{F}_{\mu\rho}\bar{F}_{\nu\sigma}h^{\rho\sigma}
                    +\frac{1}{2}(\bar{F}_{\kappa\eta}f^{\kappa\eta}-\bar{F}_{\kappa \eta}\bar{F}^\kappa~_\sigma h^{\eta\sigma})\bar{g}_{\mu\nu}-\frac{1}{4}\bar{F}^2h_{\mu\nu}\nonumber\\
                    &+&\bar{H}_{\nu}~^\rho \tilde{f}_{\mu\rho}+\bar{H}_{\mu}~^\rho \tilde{f}_{\nu\rho}-\bar{H}_{\mu\rho}\bar{H}_{\nu\sigma}h^{\rho\sigma}
                    +\frac{1}{2}(\bar{H}_{\kappa\eta}\tilde{f}^{\kappa\eta}-\bar{H}_{\kappa \eta}\bar{H}^\kappa~_\sigma h^{\eta\sigma})\bar{g}_{\mu\nu}\label{2-G}\\
                    &-&\frac{1}{4}\bar{H}^2h_{\mu\nu}.\nonumber
  \end{eqnarray}
In analyzing  the stability  of the RN-type black hole in the EMS theory with two U(1) fields,
we first consider  the  linearized   equations (\ref{per-eq1}) and (\ref{per-eq2}) because three perturbations of metric $h_{\mu\nu}$ and  vectors $a_{\mu}$ and $b_\mu$  are coupled.
These are similar to the linearized equations for the Einstein-Maxwell theory with single U(1) field $H$~\cite{Myung:2018vug}.
For  the odd-parity perturbations, one found the Zerilli-Moncrief equation which describes two physical DOF ( degrees of freedom) propagating around the RN black hole
~\cite{Zerilli:1974ai,Moncrief:1974gw}.
Also, the even-parity perturbations with two physical DOF were studied in~\cite{Moncrief:1974ng,Moncrief:1975sb}.
It turns out that the RN black hole is  stable against these perturbations.

In our case,  a massless spin-2 mode starts with  $l=2$, while two massless spin-1 mode begin with $l=1$. The EMS  theory with two U(1) provides 7(=2+2+2+1) DOF propagating around the RN-type background. We hope that the RN-type black hole is still  stable against full tensor-vector perturbations.

Now, we focus on the the linearized scalar equation (\ref{per-eq3}) which determines totally the instability of RN-type black hole  found from the EMS theory with two U(1) fields.
From  now on, we call RN-type as RN for simplicity.
 Introducing
 \begin{equation}
 \delta \varphi(t,r,\theta,\phi) =\int \sum_{lm} \varphi(r) Y_{lm}(\theta) e^{i m \phi} e^{-i\omega t} d\omega,\quad \varphi(r)=\frac{u(r)}{r}
\end{equation}
equation (\ref{per-eq3}) takes the Schr\"{o}dinger-equation with the tortoise coordinate $r_*$
\begin{equation}
\frac{d^2 u(r)}{dr^2_*} +\Big[ \omega^2-V_{\rm RN}(r)\Big] u(r) =0, \quad r_*=\int \frac{dr}{f(r)}.
\end{equation}
Here,  the potential is given by
\begin{equation} \label{RN-P}
V_{\rm RN}(r)=f(r)\Big[\frac{2M}{r^3}+\frac{l(l+1)}{r^2} -\frac{2(Q^2+P^2)}{r^4}-\frac{\alpha(P^2-Q^2)}{r^4}\Big],
\end{equation}
where the case of $P^2>Q^2$ induces the tachyonic instability depending on the coupling parameter $\alpha$. Also, the case of $P^2=Q^2$ implies no coupling effectively. We wish to delete the other case of  $P^2<Q^2$
because it induces a positive definite potential, leading to the stable RN  black hole.
In addition, the case of the same coupling leads to the last term of $-\alpha(P^2+Q^2)/r^4$, which makes no difference when comparing with the EMS theory.
This is the reason why we consider the different scalar couplings as  $e^{-\alpha \phi^2} F^2$ and $e^{\alpha \phi^2}H^2$.
\begin{figure*}[t!]
   \centering
  \includegraphics{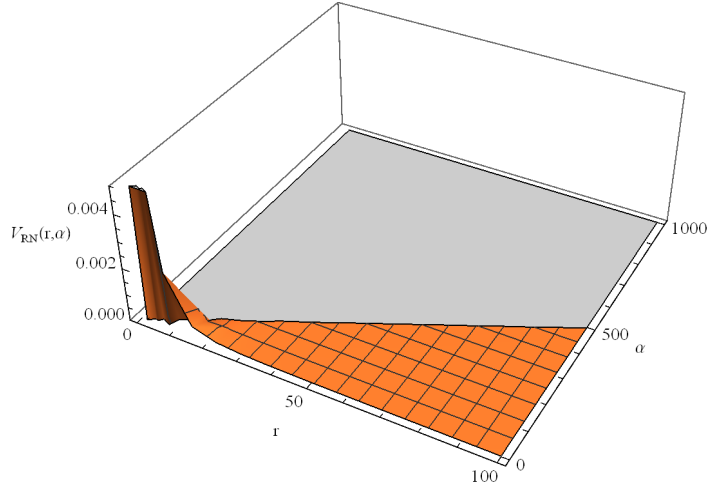}
\caption{The 3D potential $V_{\rm RN}(r,\alpha)$  as functions of $r\in [r_+=1.648,100]$ and $\alpha\in[0,1000]$ with  $p=0.7$, $q=0.3$, and $l=0$.
 The shaded region denotes negative region between $r=r_+$ and  $r= r_{out}=(p^2+q^2)+\alpha(p^2-q^2)/2$. }
\end{figure*}
In Fig. 2, we display the $(r,\alpha)$-dependent potentials for given $l=0$, $M=1$ and $p=0.7, q=0.3$ (a non-extremal RN black hole).
The negative (shaded) region appears between $r=r_+$ and $r=r_{out}=(p^2+q^2)+\alpha(p^2-q^2)/2$, whose region can be used for computing  the discrete resonant spectrum ($\{\alpha_n\}$) when employing the WKB method.
The $s(l=0)$-mode is allowed for the scalar perturbation and it is regarded as an important mode to test the stability of the RN black hole.
Hereafter,  we consider this mode only.

The sufficient condition of $\int^\infty_{r_+} dr [V_{\rm RN}(r)/f(r)]<0$ for instability~\cite{Dotti:2004sh}  leads to the bound as
\begin{equation}
\alpha> \alpha_{\rm in}(p,q)=\frac{-2(p^2+q^2)+3(1+\sqrt{1-p^2-q^2})}{p^2-q^2}, \label{sc-in}
\end{equation}
where we note that $V_{\rm RN}(r)/f(r)$ differs from $V_{\rm RN}(r)$ in Fig. 3.
On the other hand, by observing the potential (\ref{RN-P}),  the positive definite potential without negative region  could be found when imposing the bound
\begin{equation}
\alpha \le \alpha_{\rm po}(p,q)= \frac{-2(p^2+q^2)+2(1+\sqrt{1-p^2-q^2})}{p^2-q^2},\label{c-po}
\end{equation}
which  is called the sufficient condition for stability.
\begin{figure*}[t!]
   \centering
  \includegraphics{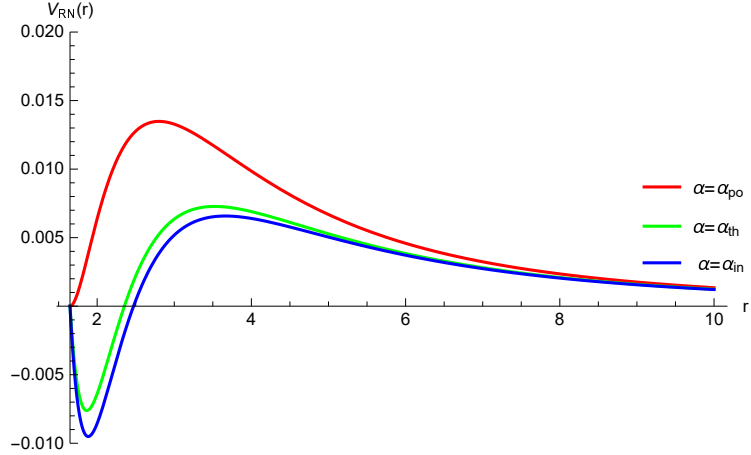}
\caption{The $\alpha$-dependent potentials as function of $r\in [r_+,10]$  with  the outer horizon radius $r_+=1.648(p=0.7,q=0.3)$ and $l=0$.
  From the top, each curve represents the potential  $V_{\rm RN}(r)$  of a scalar field for $\alpha_{\rm po}= 5.3404$ (sufficient condition for stability), $\alpha_{\rm th}=8.8646$ (threshold of instability), $\alpha_{\rm in}=9.4606$ (sufficient condition for instability), respectively. }
\end{figure*}
Fig. 3 suggests that  the threshold of instability $\alpha_{\rm th}$ is  between $\alpha_{\rm po}=5.3404$ and  $\alpha_{\rm in}=9.4606$ for $p=0.7$ and $q=0.3$.
To determine the threshold of instability $\alpha_{\rm th}$, one has to solve the second-order differential equation numerically
\begin{equation}\label{pertur-eq}
\frac{d^2u}{dr_*^2}-\Big[\Omega^2+V_{\rm RN}(r)\Big]u(r)=0,
\end{equation}
which  allows an exponentially growing mode of  $e^{\Omega t}(\omega=i\Omega) $ as  an unstable mode.
Here we choose two boundary conditions: a normalizable
solution of $u(\infty)\sim e^{-\Omega r_*}$  at infinity  and
a solution of $u(r_+)\sim \left(r-r_+\right)^{\Omega r_+}$  near the outer horizon.
We find that  the threshold ($\Omega=0$) of instability is located at  $\alpha_{\rm th}=8.86464$ for $p=0.7$ and $q=0.3$.
This implies that for given $p=0.7$ and $q=0.3$,  the RN black hole is unstable for $\alpha>\alpha_{\rm th}$ (See Fig. 10), while it is stable for  $\alpha<\alpha_{\rm th}$.
The other way of obtaining $\alpha_{\rm th}$ is to solve the  static linearized  equation directly because $\alpha_{\rm th}=\alpha^{\rm E}_{n=0}$.

We consider the  static scalar perturbed equation on the RN black hole background to identify the $n=0$, 1, 2 black holes as
\begin{equation} \label{ssclar-eq}
\frac{1}{r^2}\frac{d}{dr}\Big[r^2f(r)\frac{d\varphi(r)}{dr}\Big]-\Big[\frac{l(l+1)}{r^2}-\frac{\alpha(P^2- Q^2)}{r^4}\Big] \varphi(r)=0
\end{equation}
which describes an eigenvalue problem: for a given $l=0$, requiring an asymptotically vanishing, smooth scalar field
selects a discrete set of $n=0$, 1, 2, $\cdots$. Actually, these determine the bifurcation points (discrete resonant spectrum: $\{\alpha_n^{\rm E}\}$) numerically.
For this purpose, one may transform (\ref{ssclar-eq}) to the Legendre equation whose exact solution is given by
\begin{equation}
\varphi(r)=P_u\Big[1+\frac{2(P^2+Q^2)(r-r_+)}{r(r_+^2-Q^2-P^2)}\Big],\quad u=\frac{1}{2}\Bigg[\sqrt{1-4\alpha \Big(\frac{P^2-Q^2}{P^2+Q^2}\Big)}-1\Bigg]
\end{equation}
with the Legendre function $P_u$. Here,  we point out that there was a wrong transformation to the Legendre equation in~\cite{Astefanesei:2019pfq}.
For four parameters of $\alpha,~P(>Q),~Q,~r_+$, the function $\varphi(r)$ approaches a constant non-zero values asymptotically: $\varphi(r)\to \varphi_\infty={}_2F_1[\cdots]+{\cal O}(1/r)$ with ${}_2F_1[\cdots]$ the hypergeometric function. Finding   $\{\alpha_n^{\rm E}\}$ is equivalent to obtaining  the zeros of  ${}_2F_1[\cdots]$.
So, one has to solve the following  equation to find bifurcation points ($\{\alpha_n^{\rm E}\}$):
\begin{equation} \label{anal-br}
{}_2F_1\Big[1-u,u+1, 1, \frac{p^2+q^2}{2(p^2+q^2-1-\sqrt{1-p^2-q^2})}\Big]|_{\{p=0.7,q=0.3\}}=0.
\end{equation}
We obtain $\{\alpha_n^{\rm E}\}$ numerically and list it in Table 1.
\begin{table*}[h]
\resizebox{16cm}{!}
{\begin{tabular}{|c|c|c|c|c|c|c|c|c|c|c|c|}
\hline
$n$&0 & 1&2&3&4&5&6&7&8&9&10 \\ \hline
$\alpha_n^{\rm E}$&8.86464  & 44.6633  & 109.071  & 202.111  & 323.754  & 474.031 & 652.932  &
860.457  &1096.61 &1361.38   & 1654.78  \\ \hline
$\alpha_n$[(\ref{alphan})]&8.05054& 43.8307& 108.235& 201.264& 322.916& 473.193& 652.094&
859.619& 1095.77& 1360.54& 1653.94\\ \hline
\end{tabular}}
\caption{Results for  $\alpha_n^{\rm E}$  and $\alpha_n$ for $n=0,1,2,\cdots,10$ branches of scalarized charged black holes with $p=0.7$ and $q=0.3$. We confirm that the threshold of instability $\alpha_{\rm th}$ is given by  $\alpha_0^{\rm E}$. }
\end{table*}
We confirm a relation of $\alpha_{\rm th}=\alpha_0^{\rm E}$.
\begin{figure*}[t!]
   \centering
  \includegraphics{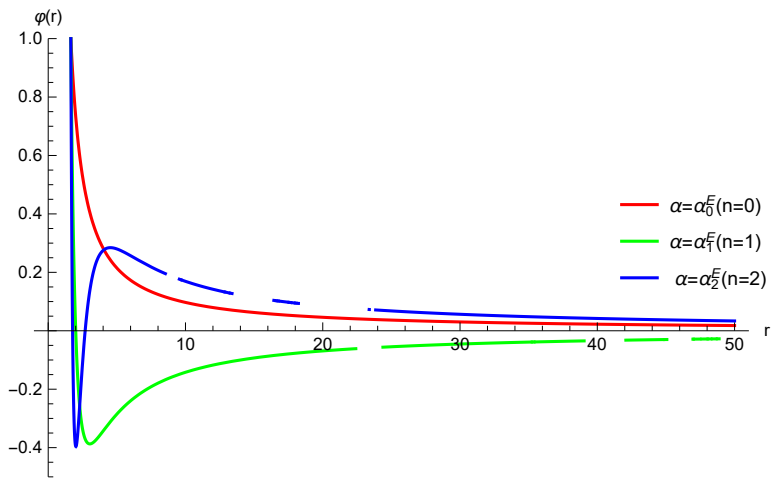}
\caption{Radial profiles of $\varphi(r)$ as function of $r\in[r_+=1.648,50]$ for the first three perturbed scalar solutions with $p=0.7$ and $q=0.3$.
These solutions are classified by the order number $n=0,1,2$ which is identified by the number of nodes (zero crossings) for  $\varphi(r)$.  }
\end{figure*}
We plot $\varphi(r)$ as a function of $r$ with three $\alpha=\alpha^{\rm E}_0,~\alpha^{\rm E}_1,~\alpha^{\rm E}_2$ whose forms can be found from Fig. 4.
These solutions are classified by the order number $n=0,~1,~2$ which is identified by the number of nodes for  $\varphi(r)$.
It is worth noting   that the $n=0$ scalar cloud without zero crossing will develop the fundamental branch of scalarized charged black hole with $\alpha\ge \alpha^{\rm E}_0$,
while the $n=1,2$ scalar clouds with zero crossings will develop the $n=1,~2$ excited branches of scalarized charged black holes with $\alpha\ge \alpha^{\rm E}_1,~\alpha^{\rm E}_2$, respectively.
\begin{figure*}[t!]
   \centering
  \includegraphics{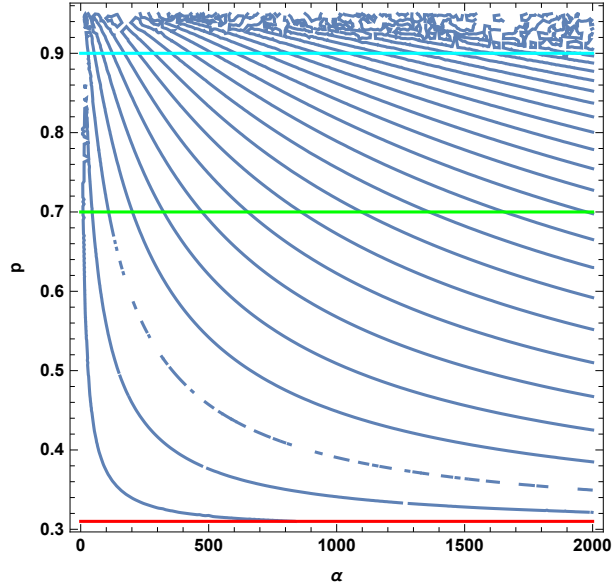}
\caption{The several curves of ${}_2F_1[\cdots]=0$ as functions of  $p\in[0.31,0.95]$  and  $\alpha\in [0,2000]$ with $q=0.3$.
The curves denote $n=0,~1,~2,\cdots$ from the left to the right. The first curve ($n=0$) represents  the boundary (existence curve) between RN black hole and scalarized charged black holes.  A green line implies  $p=0.7$ whose (eleven) crossing points  determine $\alpha_n^{\rm E}$ in Table 1. A. The red and cyan lines denote $p=0.31$ and $p=0.9$, respectively.  }
\end{figure*}

Also, we represent several curves of ${}_2F_1[\cdots]=0$ existing in $(\alpha,p)$-space (see Fig. 5)
whose crossing points with $p=0.7$ determine  $\{\alpha_n^{\rm E}\}$ in Table 1.
 For fixed $q=0.3$ and $\alpha\in[0,2000]$, the number of crossing points increase as $p$ increases.
For example, we have $\alpha_0^{\rm E}(=\alpha_{\rm th})=838.162 $ only  for $p=0.31$, while it includes 23 cases of  $\alpha_0^{\rm E}(=\alpha_{\rm th})=2.59013,\cdots,\alpha_{22}^{\rm E}=1931.45$  for $p=0.9$.
Importantly,  the first curve ($n=0$) in the left represents an existence one, which means the boundary between RN black hole and scalarized charged black holes.
In other words, this curve determines all thresholds of instability [$\alpha_{\rm th}(p,q=0.3)$] for RN black holes for any $p>0.3$.

On the other hand, it was proposed that  the spatially regular  scalar configurations (scalar clouds)  described by (\ref{pertur-eq}) with $\Omega=0$ could be investigated
analytically by making use of the standard WKB techniques~\cite{Hod:2020ljo}.
A standard second-order WKB analysis  could be applied for obtaining the bound states of the potential $V_{\rm RN}$ approximately to yield the quantization condition
\begin{equation}
\int^{r_*^{out}}_{r_*^{in}}dr_* \sqrt{-V_{\rm RN}(r_*)}=\Big(n-\frac{1}{4}\Big)\pi,\quad n=1,2,3,\cdots, \label{wkb1}
\end{equation}
where $r_*^{out}$ and $r_*^{in}$ are the radial turning points satisfying $V_{\rm RN}(r_*^{out})=V_{\rm RN}(r_*^{in})=0$.
We could express  Eq.(\ref{wkb1}) in terms of the radial coordinate  $r$ as
\begin{equation}
\int^{r_{out}}_{r_{in}}dr \frac{\sqrt{-V_{\rm RN}(r)}}{f(r)}=\Big(n-\frac{1}{4}\Big)\pi,\quad n=1,2,3,\cdots, \label{wkb2}
\end{equation}
where radial turning points $\{r_{out},r_{in}\}$ are determined by the two conditions (see Fig. 2)
\begin{equation}
1-\frac{2M}{r_{in}}+\frac{P^2+Q^2}{r^2_{in}}=0,\quad \frac{2M}{r^3_{out}}-\frac{2(P^2+Q^2)}{r^4_{out}}-\frac{\alpha(P^2-Q^2)}{r^4_{out}}=0,
\end{equation}
which admit
\begin{equation}
r_{in}=r_+,\quad r_{out}=p^2+q^2 + \frac{\alpha(p^2-q^2)}{2}.
\end{equation}
For large $\alpha(r_{out}\to \infty)$, the WKB integral (\ref{wkb2}) could be approximated by neglecting the first three terms in (\ref{RN-P}) as
\begin{equation}
\sqrt{\alpha}\int^{\infty}_{r_+} dr \sqrt{\frac{P^2-Q^2}{r^4 f(r)}}=\Big(n+\frac{3}{4}\Big)\pi,\quad  n=0,1,2,\cdots,
\end{equation}
which could be integrated analytically to yield
\begin{equation} \label{alphan}
\alpha_n(p,q)=\Big(\frac{p^2+q^2}{p^2-q^2}\Big)\Bigg[\frac{\pi \Big(n+\frac{3}{4}\Big)}{\ln\Big[\frac{\sqrt{1-p^2-q^2}}{1-\sqrt{p^2+q^2}}\Big]}\Bigg]^2,\quad  n=0,1,2,\cdots.
\end{equation}
It seems that for $p=0.7$ and $q=0.3$, $\alpha_n$  is nearly the same as the exact $\alpha_n^{\rm E}$ in Table 1.
However, one finds   that $\varphi_\infty \not=0$ for $\alpha=\{\alpha_n\}$. This implies that $\{\alpha_n\}$ determined by the WKB method does not
describe the asymptotically vanishing scalar clouds correctly. Hence,  $\{\alpha_n\}$ do not represent  bifurcation points precisely.

The infinite  $n=0,~1,~2,\cdots$ black holes with $p=0.7$ and $q=0.3$ are defined by $\alpha$-bounds of $\alpha\ge \alpha_0^{\rm E},~\alpha\ge \alpha_1^{\rm E}$, $\alpha\ge \alpha_2^{\rm E},~\cdots$, respectively.  In addition, we confirm an inequality for $p=0.7$ and $q=0.3$ as
\begin{equation}
\alpha_{\rm po}=5.3404<\alpha_0^{\rm E}=\alpha_{\rm th}=8.86464 < \alpha_{\rm in}=9.4606.
\end{equation}

\section{Scalarized charged black holes}

To obtain scalarized charged black holes through spontaneous scalarization, we  introduce  the metric and fields as~\cite{Herdeiro:2018wub}
\begin{eqnarray}\label{nansatz}
ds^2_{\rm SCBH}&=&\bar{g}_{\mu\nu}dx^\mu dx^\nu=-N(r)e^{-2\delta(r)}dt^2+\frac{dr^2}{N(r)}+r^2(d\theta^2+\sin^2\theta d\varphi^2) \nonumber \\
N(r)&=&1-\frac{2m(r)}{r},\quad\bar{\phi}= \phi(r),\quad \bar{A}_t=v_Q(r),\quad \bar{B}_t=v_P(r).
\end{eqnarray}

Substituting (\ref{nansatz}) into (\ref{equa1})-(\ref{s-equa}), one has the five equations
\begin{eqnarray}
&&-2m'(r)+e^{2\delta(r)}r^2\left(e^{-\alpha\phi(r)^2}(v_Q'(r))^2
+e^{\alpha\phi(r)^2}(v_P'(r))^2\right) \nonumber \\
&&\quad+r[r-2m(r)](\phi'(r))^2=0, \label{neom1}\\
&&\delta'(r)+r(\phi'(r))^2=0, \label{neom2}\\
&&v_Q'(r)\Big(2+r\delta'(r)-2r\alpha\phi(r)\phi'(r)\Big)+r v_Q''(r)=0, \label{neom3}\\
&&v_P'(r)\Big(2+r\delta'(r)+2r\alpha\phi(r)\phi'(r)\Big)+r v_P''(r)=0, \label{neom4}\\
&&e^{2\delta(r)}r^2\alpha\phi(r)\left(e^{\alpha\phi(r)^2}(v_P'(r))^2
-e^{-\alpha\phi(r)^2}(v_Q'(r))^2\right)+r[r-2m(r)]\phi''(r)\nonumber\\
&&-\Big(m(r)[2-2r\delta'(r)]
+r[-2+r\delta'(r)+2m'(r)]\Big)\phi'(r)=0, \label{neom5}
\end{eqnarray}
where the prime ($'$) denotes differentiation with respect to its argument.

Accepting the existence of a horizon located at $r=r_+$,  one finds an
approximate solution to equations (\ref{neom1})-(\ref{neom5}) in the near-horizon
\begin{eqnarray}
&&m(r)=\frac{r_+}{2}+m_1(r-r_+)+\cdots,\quad
\delta(r)=\delta_0+\delta_1(r-r_+)+\cdots,\label{aps-1}\\
&&v_{Q}(r)=v_{Q1}(r-r_+)+\cdots,\quad \quad
v_{P}(r)=v_{P1}(r-r_+)+\cdots,\label{aps-2}\\
&&\phi(r)=\phi_0+\phi_1(r-r_+)+\cdots,\label{aps-3}
\end{eqnarray}
where the five coefficients are given by
\begin{eqnarray}\label{ncoef}
&&m_1=\frac{e^{-\alpha\phi_0^2}P^2+e^{\alpha\phi_0^2}Q^2}{2r_+^2},\quad
\delta_1=-r_+\phi_1^2,\nonumber\\
&&\phi_1=\frac{\alpha \phi_0(P^2-e^{2\alpha\phi_0^2}Q^2)}{r_+(P^2+e^{2\alpha\phi_0^2}Q^2-e^{\alpha\phi_0^2}r_+^2)},\nonumber\\
&&v_{Q1}=-\frac{e^{-\delta_0+\alpha\phi_0^2}Q}{r_+^2}, \quad v_{P1}=-\frac{e^{-\delta_0-\alpha\phi_0^2}P}{r_+^2}.
\end{eqnarray}
Here, two important parameters of $\phi_0=\phi(r_+,\alpha)$ (See Fig.6) and $\delta_0=\delta(r_+,\alpha)$ are  determined when matching with
an asymptotically flat solution in the far-region
\begin{eqnarray}\label{ncoef}
&&m(r)=M-\frac{P^2+Q^2+Q_s^2}{2r}+\cdots,\quad
\delta(r)=\frac{Q_s^2}{2r^2}+\cdots,\nonumber\\
&&v_{P}(r)=\Phi_P+\frac{P}{r}+\cdots,\qquad \qquad \quad
v_{Q}(r)=\Phi_Q+\frac{Q}{r}+\cdots,\nonumber\\
&&\phi(r)=\frac{Q_s}{r}+\cdots,
\end{eqnarray}
where  $Q_s$, $\Phi_Q$ and $\Phi_P$ denote the scalar charge, and the electrostatic potentials at infinity, in addition to the ADM mass $M$, and the electric charges $Q$ and $P$.
\begin{figure*}[t!]
\centering
\includegraphics{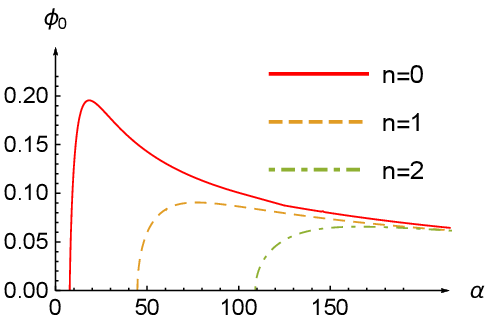}
\hfill%
\includegraphics{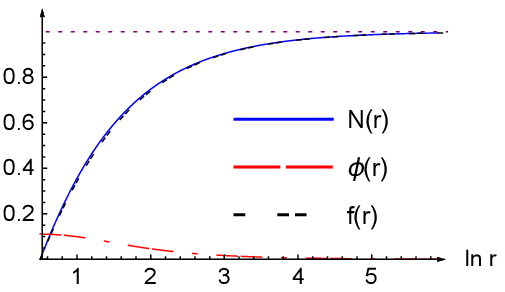}
\hfill%
\includegraphics{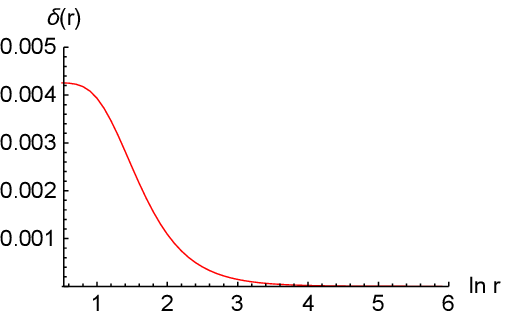}
\caption{(Left) The scalar field $\phi_0=\phi(r_+)$ at the horizon as function of $\alpha$.
The $n=0$ fundamental branch starts from the first bifurcation point at $\alpha^{\rm E}_0=8.864$, while $n=1, 2$
excited branches start from  $\alpha^{\rm E}_1=44.663$ and $\alpha^{\rm E}_2=109.071$.
(Middle and Right) Graphs of a scalarized charged black hole with $\alpha=68.45$,
$\phi_0=0.111$,  and $\delta_0=0.0043$ in the $n=0$ branch with $P=0.7$ and $Q=0.3$.
Here $f(r)$ represents the metric function for the RN black hole with $\phi_{\rm RN}(r)=0$ and $\delta_{\rm RN}(r)=0$.
We plot all figures in terms of $\ln r$ and thus, the horizon is always located at $\ln r=\ln r_+=-0.153$.}
\end{figure*}

At this stage, we wish to comment that there is no constraint on $P$ and $Q$ in constructing scalarized charged black holes.

As an explicit scalarized charged black hole solution with $P=0.7$ and $Q=0.3$,
we show a numerical black hole solution with $\alpha=68.45$ in the $n=0$ fundamental  branch of $\alpha\ge 8.864$ in Figs. 6 and 7.
However, we need hundreds of numerical solutions depending $\alpha$ for each branch to perform the stability of scalarized charged black holes.
\begin{figure*}[t!]
\centering
\includegraphics{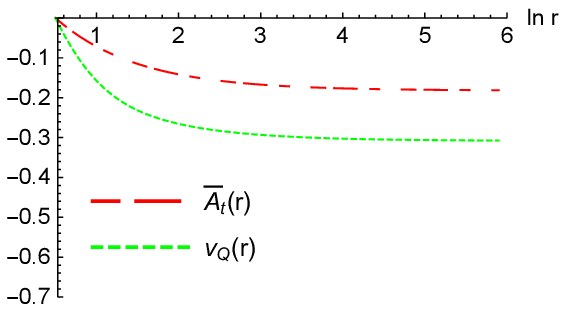}
\hfill%
\includegraphics{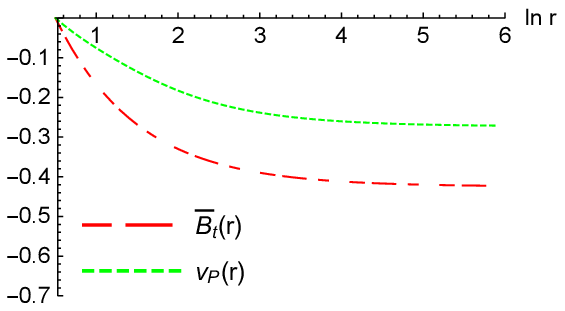}
\caption{ $\bar{A}_{t}(r)$ and $\bar{B}_{t}(r)$ represent two vector potentials of RN black holes.
$v_{Q}(r)$ and $v_{P}(r)$ represent vector potentials of
scalarized charged black hole with $\alpha=68.45$, $\Phi_P=-0.424$ and $\Phi_Q=-0.182$.}
\end{figure*}

\section{Stability of scalarized charged black holes}
The stability of scalarized charged black holes is an important question because it determines their viability in representing realistic
astrophysical configurations.
We prefer to  introduce the radial perturbations around the scalarized black holes as
\begin{eqnarray}
&&ds_{\rm rad}^2=-N(r)e^{-2\delta(r)}(1+\epsilon H_0)dt^2+\frac{dr^2}{N(r)(1+\epsilon H_1)}
+r^2(d\theta^2+\sin^2\theta d\psi^2),\nonumber\\
&&H_{rt}(t,r)=v'_{P}(r)+\epsilon\delta v_P(t,r),\quad
F_{rt}(t,r)=v'_{Q}(r)+\epsilon\delta v_Q(t,r),\nonumber\\
&&\phi(t,r)=\phi(r)+\delta\tilde{\phi}(t,r),
\end{eqnarray}
where $N(r)$, $\delta(r)$, $\phi(r)$, $v_{P}(r)$ and $v_{Q}(r)$ represent a scalarized
charged black hole background, while
$H_0(t,r)$, $H_1(t,r)$, $\delta\tilde{\phi}(t,r)$, $\delta v_{P}(t,r)$ and $\delta v_{Q}(t,r)$
denote five perturbed fields around the scalarized
black hole background. From now on, we confine ourselves to analyzing the $l=0$(s-mode)
propagation, implying that higher angular momentum modes $(l\neq0)$ are excluded. In this
case, all perturbed fields except the perturbed scalar field may belong to redundant fields.
After applying decoupling process to linearized equations, one may find a linearized scalar equation.

Considering the separation of variables
\begin{eqnarray}
\delta\tilde{\phi}(t,r)=\frac{\tilde{\varphi}(r)e^{\Omega t}}{r},
\end{eqnarray}
we obtain the Schr\"{o}dinger-type equation for an $s$-mode scalar perturbation
\begin{eqnarray}
\frac{d^2\tilde{\varphi}(r)}{dr_*^2}-\Big[\Omega^2+V(r,\alpha)\Big]\tilde{\varphi}(r)=0,
\end{eqnarray}
with $r_*$ is the tortoise coordinate defined by
\begin{eqnarray}
\frac{dr_*}{dr}=\frac{e^{\delta(r)}}{N(r)}.
\end{eqnarray}
Here, its potential reads to be
\begin{eqnarray} \label{sc-poten}
V(r,\alpha)=\frac{N}{e^{2\delta}r^2}\Big[(1-N-2r^2\phi'^2)
&+&\frac{e^{-\alpha\phi^2}P^2[-\alpha-1+2(-\alpha\phi+r\phi')^2]}{r^2}\nonumber\\
&+&\frac{e^{\alpha\phi^2}Q^2[\alpha-1+2(\alpha\phi+r\phi')^2]}{r^2}\Big]
\end{eqnarray}
whose limit of $Q^2\to 0$ recovers the potential $U_{\Omega}(-\alpha)$ for the EMS theory in~\cite{Herdeiro:2018wub}.
\begin{figure*}[t!]
   \centering
  \includegraphics{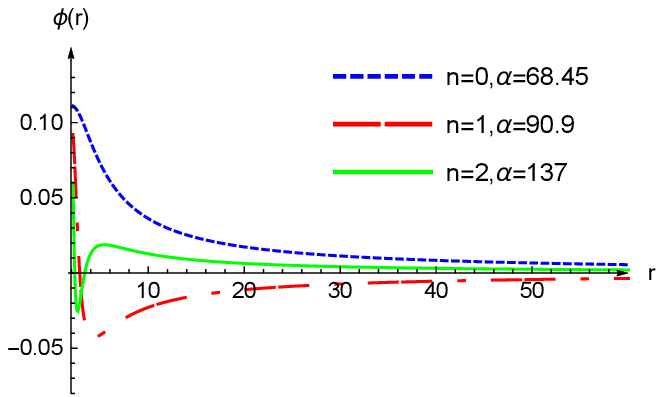}
  \hfill%
\includegraphics{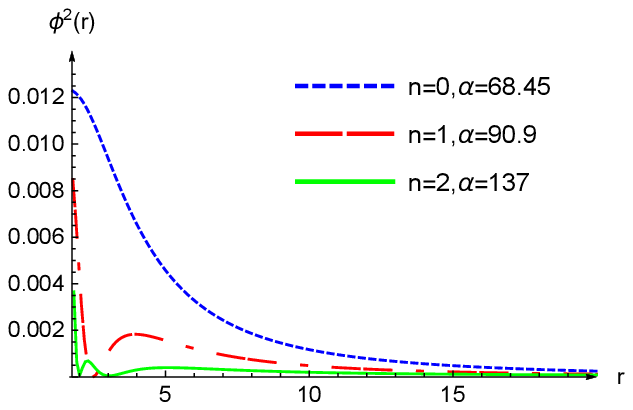}
\caption{(Left) Scalar solution  $\phi(r)$  and (Right) its square $\phi^2(r)$ as functions of $r\in[r_+=1.648,50]$ for the first three branches with $P=0.7$ and $Q=0.3$.
Here, we choose  $\phi_0 = 0.111$,~  $Q_s=0.3272$ for  $n=0$ branch ; $\phi_0$ = 0.0921,~  $Q_s=-0.2097$ for $n=1$ branch; $\phi_0 = 0.0606$,~  $Q_s=0.1222$, for $n=2$ branch.}
\end{figure*}

Before we proceed, we wish to analyze the potential $V(r,\alpha)$ carefully because it is a compact one.
First of all, we observe that $V(r,\alpha)$ reduces to $V_{\rm RN}(r)$ in (\ref{RN-P}) when imposing $\phi=\delta=0[N(r)\to f(r)]$.
This implies that `$-\alpha P^2/r^2$' in the second term  contributes to a negatively large potential in the near-horizon  as in the RN case (see Fig. 3), while the first and last terms make  positively small contributions to the potential. Importantly, `$2(-\alpha \phi+r\phi')^2P^2/r^2$' in the second term plays the role of  making   small positive region in the near-horizon as $n$ increases. As is shown in Fig. 8 (similar to Fig. 4), the number of scalar-node increases as $n$ increases, which implies that the positive (negative) region of $\phi^2$ ($V(r,\alpha)$) decreases (increases) in the near-horizon. This may explain that the $n=0(\alpha\ge 8.864)$ black hole is stable against the $s$-mode scalar perturbation, whereas the  $n=1(\alpha\ge 44.67),(n=2(\alpha\ge 109.071),~3(\alpha\ge 202.111),~4(\alpha\ge 323.754),~\cdots$ excited black holes may be  unstable.
\begin{figure*}[t!]
\centering
\includegraphics{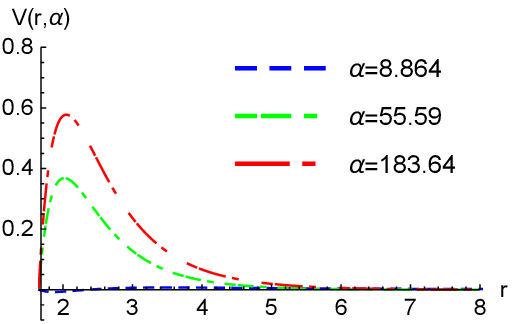}
\hfill%
\includegraphics{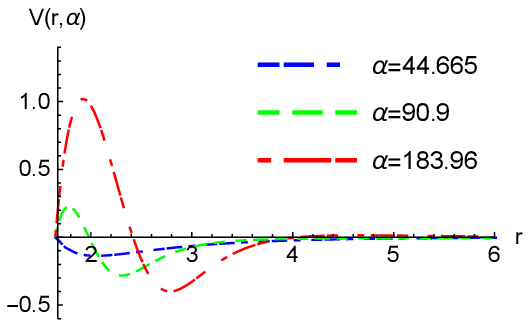}
\hfill%
\includegraphics{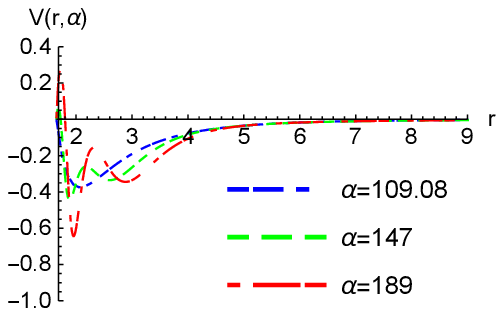}
\caption{ Scalar potentials $V(r,\alpha)$ around $n=0$ (Left: $\alpha\ge8.864$), 1 (Middle: $\alpha\ge 44.663$),
2 (Right: $\alpha\ge109.071$) black holes in the infinite branches. The positive barriers in the near-horizon become smaller and smaller as $n$ increases.  }
\end{figure*}

The conclusions about the stability of the scalarized charged black holes with respect to radial perturbations will be reached by examining
the qualitative behavior of the potential $V(r,\alpha)$ as well as by obtaining explicitly exponentially growing (unstable) modes for $s$-mode scalar  perturbation.
We display three scalar potentials $V(r,\alpha)$ in (Left) Fig. 9 for $l=0(s$-mode) scalar around the $n=0$ black hole, showing positive definite.
This implies that the $n=0$ black hole is stable against the $s$-mode of perturbed scalar. We confirm its stability  by noting negative $\Omega$ in Fig. 10.
 We observe from Fig. 9 that $\int_{r_+}^\infty dr[e^\delta V(r,\alpha)/N]<0$ (sufficient condition for instability~\cite{Dotti:2004sh}) for the $n=1,~2$ black holes.
This suggests that the $n=1,~2$ black holes are unstable against the $s$-mode scalar perturbation.
 Obviously, their instability are found from Fig. 10 in accordance with the existence of unstable modes because all ($\Omega$)  are positive.
This is consistent with the results for the EMS theory with exponential coupling~\cite{Myung:2018jvi} and quadratic coupling~\cite{Myung:2019oua}, and for the EMCS theory with exponential and quadratic couplings~\cite{Zou:2020zxq}. The stability for a quartic coupling in the EMS theory was recently announced by considering full perturbations~\cite{Blazquez-Salcedo:2020jee}.

\begin{figure*}[t!]
\centering
\includegraphics{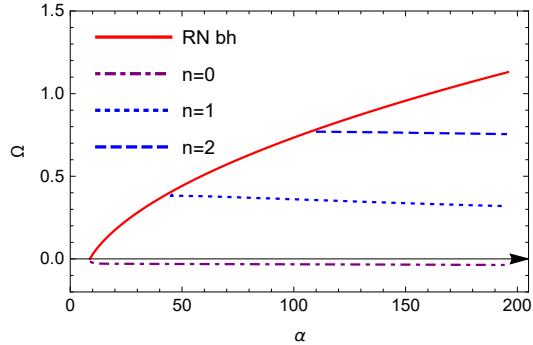}
\caption{Plots of $\Omega$ as functions of $\alpha$ for $l=0$-scalar mode around the $n=0(\alpha\ge8.864)$,
$1(\alpha\ge44.663)$, $2(\alpha\ge109.071)$ black holes with $P=0.7$ and $Q=0.3$. The positive $\Omega$ for $n=1$, 2 black holes implies unstable
black holes, while the negative $\Omega$ for the $n=0$ black hole shows a stable black hole. A red curve started at $\alpha=8.864(\Omega=0)$
denotes the positive $\Omega$, indicating the unstable RN black hole for $\alpha>8.864$.}
\end{figure*}

\section{Discussions }
First of all, let us compare a fixed scalar in the dilatonic black hole with scalar hairs in the scalarized charged black holes.
Fig. 1 implies that the dilaton is a fixed scalar  whose value on the horizon is fixed by the U(1) charges $ Q$ and $P$ and it is given independently by $\phi_\infty=0$ at infinity.
A scalar hair whose value on the horizon is fixed by  two U(1) charges $Q,~P$ and $\alpha$, and it  is asymptotically zero.
Hence, the fixed scalar is similar to the scalar hair.

Now, we would like to mention the stability of scalarized charged black holes.
Firstly, we note that the RN black hole is unstable for $\alpha>\alpha_{\rm th}$ (See Fig. 10), while it is stable for  $\alpha<\alpha_{\rm th}$. Here, $\alpha_{\rm th}$ denotes the threshold of instability for RN black hole as well as it indicates the boundary between  RN and $n=0$ scalarized charged black holes.
On the region ($\alpha \ge 8.864$) of unstable RN black holes, we could obtain the fundamental  $n=0(\alpha\ge 8.864)$ black hole, and  $n=1(\alpha\ge 44.67),~(n=2(\alpha\ge 109.071),~3(\alpha\ge 202.111),~4(\alpha\ge 323.754),~\cdots$ excited black holes inspired by the onset of spontaneous scalarization.

The stability analysis of scalarized charged black holes is an  important matter because it determines their viability in representing realistic astrophysical configurations.
Also, it is not an easy task  since one needs hundreds of numerical solutions depending $\alpha$ for each branch to perform the stability of scalarized charged black holes.
It seems to be  difficult for them to become stable ones because their defined areas correspond to region of unstable RN black holes without scalar hair (by making large negative region in the potential).
Fortunately, one may have a stable black hole in the fundamental branch because there exists a positively scalar hair contribution of `$2(-\alpha \phi+r\phi')^2P^2/r^2$' to the potential (\ref{sc-poten}).
It turns out that the $n=0$  black hole is stable against the $s(l=0)$-mode scalar perturbation, whereas the $n=1,2$ excited black holes are unstable. This is consistent with the results for the EMS theory with exponential coupling~\cite{Myung:2018jvi} and quadratic coupling~\cite{Myung:2019oua}.

On the other hand, the stability analysis of scalarized charged  black holes can answer to  whether they could be the endpoints of tachyonic instability of RN black holes without scalar hair. Since the $n=0$ scalarized charged black hole is  stable, this is  regarded as an endpoint of  the unstable RN black hole.

 \vspace{1cm}

{\bf Acknowledgments}

 This work was supported by the National Research Foundation of Korea (NRF) grant funded by the Korea government (MOE)
 (No. NRF-2017R1A2B4002057).
 
\end{document}